**A Monte Carlo study of thermal conductivity in black phosphorene**

**L. O. Vilyaev[1*], D. V. Zavyalov[1], V. I. Konchenkov[1,2]**

[1]Volgograd State Technical University, Russia, 400005, Volgograd, Lenin av., 28

[2]Volgograd State Social-Pedagogical University, Russia, 400005, Volgograd, Lenin av., 27

sinegordon@gmail.com

**Abstract**. In this work the electronic thermal conductivity of black phosphorene is studied. Using the Monte Carlo method together with the Green-Kubo relations, a statistical modelling of the motion and scattering of electrons on acoustic and optical phonons is performed to estimate the electronic thermal conductivity. The scattering probability is expressed analytically. The obtained results show a pronounced anisotropy of the thermal conductivity: the thermal conductivity values along the armchair direction are an order of magnitude larger than those along the zigzag direction, in agreement with the experimental data for related materials. In addition, the modelling allowed a more rigorous justification of some results of previous theoretical works on the anisotropic thermal conductivity of black phosphorene.

## Introduction

Phosphorene is a two-dimensional material obtained by mechanical or liquid-phase exfoliation of bulk black phosphorus [1, 2]. First experimentally obtained in 2014, it has attracted attention because of its direct band gap, which depends on the number of layers and on external influences [3, 4], as well as its high mobility of the carriers.

The crystal lattice of black phosphorene has a puckered structure, which leads to a pronounced anisotropy of its physical properties. Depending on the crystallographic direction (armchair or zigzag), the electronic, mechanical and phonon properties of the material differ significantly [5]. In particular, the thermal conductivity of phosphorene exhibits a strong dependence on the propagation direction of the heat flux, as well as on the number of layers, defects and temperature [6, 7], which makes it an interesting object for fundamental and applied research. Most of the works are devoted to the phonon-mediated part of phosphorene thermal conductivity [7-10]. It has been shown that the phonon thermal conductivity along the $ZZ$ (zigzag) direction is several times larger than the thermal conductivity in the $AC$ (armchair) direction [11, 12], although the absolute thermal conductivity values obtained in different studies differ considerably, which is due to the use of various methods (the classical molecular dynamics method [8, 13], the method of solving the Boltzmann kinetic equation for phonons [7, 9-11]).

This work is a further development of the study [14, 15], in which the estimates of the surface electronic thermal conductivity coefficient of phosphorene along two crystallographic directions, $AC$ (armchair) and $ZZ$ (zigzag), were obtained (Fig. 1). It was shown in [14, 15] that, unlike the phonon part, the electronic thermal conductivity along the $2AC$ direction is an order of magnitude larger than the electronic thermal conductivity along the $ZZ$ direction, owing to the substantial difference between the values of the diagonal components of the effective mass tensor of black phosphorene [16]. It was also shown in [14, 15] that, from the measurements of the angle $\Phi$ between the heat flux direction and the temperature gradient, the electronic contribution to the heat flux can be numerically isolated. The feature of [14, 15] is the use of the constant relaxation-time approximation $\tau$, which does not explicitly take into account the microscopic mechanisms of charge carrier scattering on the crystal lattice inhomogeneities. This article attempts to provide a more rigorous justification of the results obtained in [14, 15]. To this end, a model with explicit electron-phonon scattering is developed, in which the carrier dynamics in a single-layer phosphorene is calculated by the Monte Carlo method with the Green-Kubo approach, while the scattering probabilities of charge carriers on crystal lattice inhomogeneities are computed analytically.

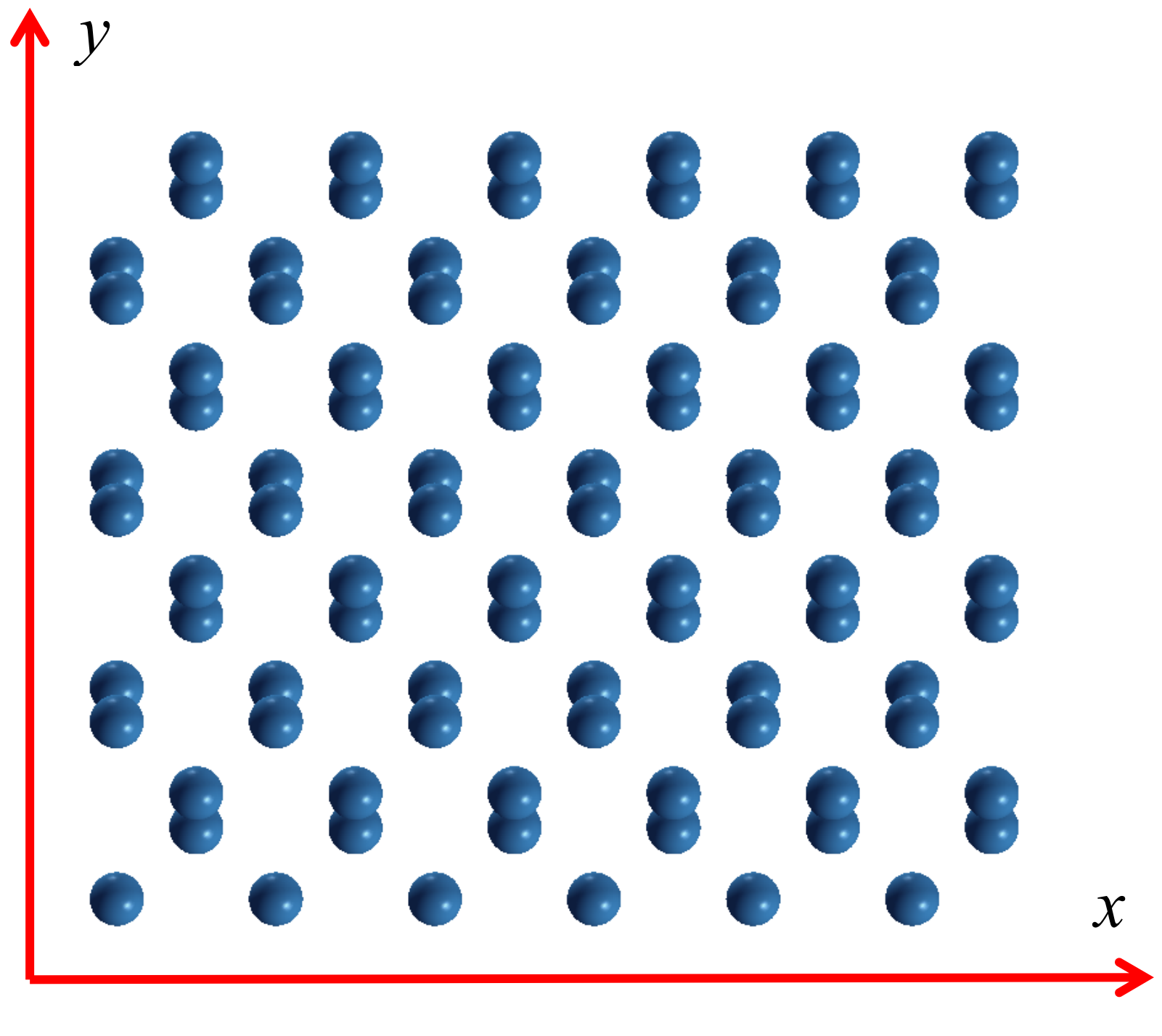

**Fig. 1.1** Geometry of the problem: orientation of the coordinate system relative to the crystallographic axes of the black phosphorene sample

The article is structured as follows: Section 1 describes the equilibrium correlation algorithm for the calculation of the two-dimensional thermal conductivity of phosphorene by the Green-Kubo method; in Section 2 the analytical derivation of the collision integral is given; Section 3 reports the results and their discussion.

## 1. Algorithm for calculating thermal conductivity by the Green–Kubo method

To compute the thermal conductivity coefficients, the approach based on the Green-Kubo method [17] was used. In contrast to the non-equilibrium method with a prescribed temperature gradient (the so-called reservoir method; see, e.g., [18]), the Green-Kubo method uses fluctuations of the heat current in a system maintained at a uniform temperature. Each component of the thermal conductivity tensor $\kappa_{\alpha\beta}$ (where $\alpha, \beta = x, y$) is obtained from the correlation of the corresponding projections of the heat flux vector:

$$\kappa_{\alpha\beta} = \frac{1}{V k_B T^2} \int_0^{\infty} \langle J_{q,\alpha}(t) J_{q,\beta}(0) \rangle dt, \tag{1}$$

where $J_{q,\alpha}(t)$ is the component of the heat flux vector along the $\alpha$ axis at the time $t$, $k_B$ is the Boltzmann constant, $T$ is the absolute temperature and $V$ is the volume of the simulated system.

1) Implementation of the Monte Carlo Method

A rectangular two-dimensional region of size $L_x \times L_y$ is considered. An axis $s$ is chosen along which the heat and particle fluxes will be calculated. The system is simulated at the equilibrium temperature $T_0$.

The carrier energy is given by the anisotropic dispersion relation with parameter $\Delta$ and effective masses $m_x$, $m_y$ [19]:

$$\varepsilon(\mathbf{p}) = \sqrt{\Delta \frac{p_x^2}{m_x} + \left( \Delta + \frac{p_y^2}{2m_y} \right)^2}. \tag{1}$$

The components of the group velocity are obtained as derivatives of the energy with respect to the momentum:

$$v_x = p_x \frac{\Delta}{m_x \varepsilon(\mathbf{p})},$$

$$v_y = p_y \frac{\Delta + \frac{p_y^2}{2m_y}}{m_y \varepsilon(\mathbf{p})}.$$

The thermal energy of the carrier is measured from the edge of the conduction band:

$$\epsilon(\mathbf{p}) = \varepsilon(\mathbf{p}) - \Delta.$$

The particle ensemble is initialized as follows. For each of the $N$ independent trajectories the initial coordinate is chosen uniformly in the region $L_x \times L_y$. The momentum is sampled from the equilibrium distribution at temperature $T_0$.

The programme has two operating modes that differ in the statistical distribution of the charge carriers.

In the Boltzmann-statistics mode the distribution

$$f_B(\mathbf{p}, T_0) \propto \exp\left(-\frac{\varepsilon(\mathbf{p})}{k_B T_0}\right)$$

is used.

In the Fermi–Dirac-statistics mode the following distribution

$$f_{FD}(\mathbf{p}, \mu, T_0) = \frac{1}{\exp\left(\frac{\varepsilon(\mathbf{p}) - \mu}{k_B T_0}\right) + 1}$$

is used.

The chemical potential $\mu$ is determined from the specified surface concentration $n_{2D}$ by numerical integration over momentum space, with $\varepsilon(\mathbf{p})$ given by Eq. (2):(1)

$$n_{2D} = \frac{g_s}{(2\pi\hbar)^2} \int f_{FD}(\varepsilon(\mathbf{p}), \mu, T_0) d^2\mathbf{p},$$

where $g_s = 2$ is the spin degeneracy.

Particle motion is calculated with a constant time step $\delta t = 10^{-16}$ s $=$ 0.1 fs. The coordinate is updated by the explicit scheme:

$$r_{n+1} = r_n + v(p_n)\delta t.$$

When a particle crosses a boundary its coordinate is returned periodically:

$$x < 0 \to x + L_x, \qquad x \geq L_x \to x - L_x;$$

$$y < 0 \to y + L_y, \qquad y \geq L_y \to y - L_y.$$

For each scattering mechanism $k$ a random free path $\lambda_k = -\ln \xi$ is stored, where $\xi$ is a random number uniformly distributed on the interval $(0, 1)$. At each time step it is decreased by $W_k \delta t$:

$$\lambda_{k+1} = \lambda_k - W_k(\mathbf{p}, s)\delta t. \qquad (2)$$

As soon as $\lambda_k < 0$, a scattering attempt is registered. In the Boltzmann-statistics mode the scattering event is always accepted. In the Fermi–Dirac-statistics mode the occupation of the final state is additionally taken into account, and the scattering is accepted with the probability:

$$P_{accept} = 1 - f_{FD}(\varepsilon_{final}, \mu, T_0).$$

Thus, occupied final states suppress scattering, which implements the Pauli exclusion principle.

2) Calculation of the macroscopic fluxes

After the start of the Monte Carlo algorithm, 20 ns of model time is allowed for equilibration of the particle ensemble under consideration, after which the fluxes are calculated over a time 80 ns every $\delta t_c = 10$ fs. The ratio between the ensemble preparation time and the actual flux simulation time is chosen empirically. Along the chosen axis $s$, where $s = x$ or $s = y$, two macroscopic fluxes are calculated at each step – the heat and particle fluxes:

$$q_s(t) = (\varepsilon(t) - \Delta)v_s(t),$$
$$j_s(t) = v_s(t).$$

For the ensemble of $N$ model trajectories the total time series are accumulated:

$$Q_s(t) = \sum_{i=1}^{N} q_{s,i}(t),$$
$$J_s(t) = \sum_{i=1}^{N} j_{s,i}(t).$$

To avoid storing the coordinates and momenta of all particles at each step, only the values of the total fluxes are saved.

3) Computation of correlation functions

The autocorrelation of the heat flux $C_{QQ}$, of the particle flux $C_{JJ}$, and the cross-correlations $C_{QJ}$ are computed from the equilibrium time series. Since $N$ independent model trajectories are used in the code, the correlation of the total mean flux is scaled so that it corresponds to the single-particle correlation.

$$C_{QQ}(\tau) = \langle q_s(t+\tau)\, q_s(t)\rangle,$$
$$C_{QJ}(\tau) = \langle q_s(t+\tau) j_s(t)\rangle,$$
$$C_{JJ}(\tau) = \langle j_s(t+\tau) j_s(t)\rangle.$$

In discrete form, for a sliding window of length $m$ time samples ($\tau = m\delta t_c$):

$$C_{QQ}[m] = \frac{N}{N_t - m} \sum_{n=0}^{N_t-m-1} \left(\frac{Q[n+m]}{N}\right)\left(\frac{Q[n]}{N}\right).$$

Here $N_t$ is the total number of time samples in each model trajectory. $C_{QJ}$ and $C_{JJ}$ are calculated similarly. For each value of $m = 1, 2, \ldots, 1000$, the correlations are averaged over all admissible initial time moments within the considered time series. Thus, the maximum interval between two moments of time at which the correlation is calculated equals $\tau_{max} = 1000\delta t_c = 10$ пс.

The calculations show that the autocorrelation of the heat flux $C_{QQ}(\tau)$ decreases rapidly with the growth of $\tau$; its main amplitude decays after 0.1– 0.2 ps of model time. The root-mean-square amplitude $C_{QQ}(\tau)$ over the interval $\tau = 5$– 10 ps is of the order of $10^{-4}$ of the amplitude $C_{QQ}(0)$, so the contribution of the autocorrelation function to the Green-Kubo integrals over this range is very small.

4) Computation of the Green-Kubo integrals and the thermal conductivity coefficient

The correlation functions are integrated over time by the trapezoidal rule. For each upper limit $\tau$ the following integrals are obtained:

$$I_{QQ}(\tau) = \int_0^{\tau} C_{QQ}(t)dt,$$

$$I_{QJ}(\tau) = \int_0^{\tau} C_{QJ}(t)dt,$$

$$I_{JJ}(\tau) = \int_0^{\tau} C_{JJ}(t)dt.$$

The conventional estimate of the two-dimensional thermal conductivity:

$$\kappa_{2D}(\tau) = \frac{n_{2D}}{k_B T_0^2} I_{QQ}(\tau).$$

For an electron gas the heat flux is coupled to the particle flux. To obtain the thermal conductivity under open-circuit electrical conditions $J = 0$, the component of the heat flux that is correlated with the particle flux is subtracted [20]:

$$\kappa_{(2D,J=0)}(\tau) = \frac{n_{2D}}{k_B T_0^2}\left(I_{QQ}(\tau) - \frac{I_{QJ}^2(\tau)}{I_{JJ}(\tau)}\right).$$

To convert the result into a volume thermal conductivity it is necessary to divide the two-dimensional value by the thickness $d$ of the sample:

$$\kappa_{(J=0)}(\tau) = \frac{n_{2D}}{k_B T_0^2 d}\left(I_{QQ}(\tau) - \frac{I_{QJ}^2(\tau)}{I_{JJ}(\tau)}\right).$$

5) Block estimate of statistical stability

For a rough estimate of the statistical uncertainty the time series of duration 80 ns is split into $N_b = 8$ sequential segments – blocks. For each block the correlations, integrals and the values of the thermal conductivity coefficient $\kappa_b$ are computed independently. In the computation of the correlations, pairs of time samples belonging to different temporal blocks are discarded. For the maximum correlation time $\tau_{max} = 10$ ps the number of admissible pairs inside one block remains of the order $10^6$. Calculations show that the plots of the Green-Kubo integrals versus time $I_{QQ}(\tau)$, $I_{QJ}(\tau)$, $I_{JJ}(\tau)$ reach a plateau at the moment $\tau_{corr.loss} = 0.5\text{–}1.0$ ps, which can be regarded as the physical correlation decay time. Thus, the ratio of the block length $\tau_b = 10$ ps to the chosen maximum correlation time $\tau_{max}$ is $10^3$, and the ratio $\tau_b/\tau_{corr.loss}$ is $10^4$. The standard deviation of the thermal conductivity coefficient $\kappa$ is given by:

$$\sigma_\kappa \equiv SD_{block} = \sqrt{\frac{1}{N_b - 1}\sum_b (\kappa_b - \bar{\kappa})^2}.$$

In the present derivation $\sigma_\kappa$ is the standard deviation of $N_b$ conductivity values computed from the selected time blocks, rather than the error of the mean. If required, the error of the mean can be obtained as:

$$SEM = \frac{\sigma_\kappa}{\sqrt{N_b}}.$$

The described algorithm is implemented as a C++ programme and is available at the link [21].

6) Analysis of the convergence of the thermal conductivity algorithm

Fig. 2 shows the dependence of the electronic thermal conductivity coefficient $4\kappa_{2D}(\tau)$, as well as of the electronic thermal conductivity coefficient $\kappa_{2D,J=0}(\tau)$ under open-circuit conditions, along the $OX$ (zigzag) and the $OY$ (armchair) axes for one of the time blocks. The plots demonstrate that the correlations decay well before $\tau_{max} = 10$ ps. The plateau of the $\kappa_{2D}(\tau)$ dependence is established by the time $\tau_{corr.loss} = 0.5\text{–}1.0$ ps. On the interval $\tau = 1\text{–}5$ ps the relative variations of the Green- Kubo integrals (and, correspondingly, of the thermal conductivity coefficient) amount to a few percent. In the current implementation of the algorithm there is no automatic plateau detector, and the final thermal conductivity coefficient value $\kappa_{2D}$ is taken at the last point corresponding to the correlation time $\tau = \tau_{max} = 10$ ps. The deviation of this value from the mean values over the plateau on the interval $\tau = 1\text{–}5$ ps is approximately 1–4%, i.e. less than $\sigma_\kappa$ over the time blocks. The integral $I_{JJ}(\tau)$ remains positive and stable, therefore the open-circuit correction $J = 0$ does not introduce any singularities on the $\kappa_{2D,J=0}(\tau)$ plots.

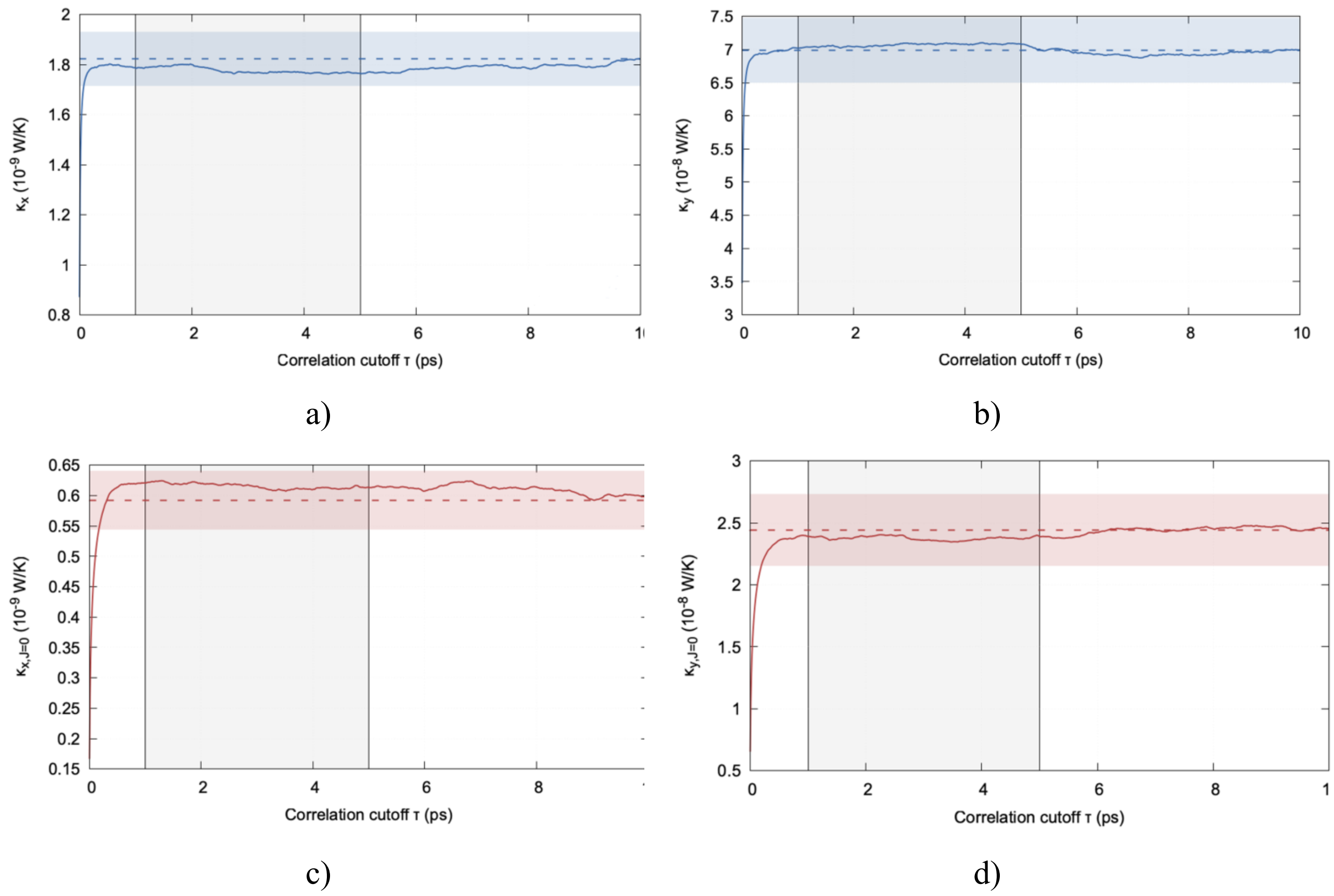

**Fig. 2.2** Dependence of the two-dimensional electronic thermal conductivity coefficient values on the correlation time $\tau$:

a), b) - thermal conductivity coefficient $\kappa_{2D}(\tau)$ along the zigzag and armchair directions, respectively;

c), d) - thermal conductivity coefficient $\kappa_{2D,J=0}(\tau)$ along the zigzag and armchair directions, respectively.

The solid line shows the thermal conductivity dependence computed for one of the time blocks of duration 10 ps. The dashed line is the mean value of the thermal conductivity coefficient $\bar{\kappa}$, averaged over $N_b = 8$ independent time blocks. The horizontal band is the region of $\bar{\kappa} \pm \sigma_\kappa$ values; the vertical boundaries mark the window 1– 5 ps

**2. Probability of electron scattering by phonons**

For the described algorithm, it is necessary to obtain the formulas for the electron scattering probability by optical and acoustic phonons per unit time (3).(2)

The probability per unit time of charge carrier scattering on acoustic phonons with a transition from a state with quasi-momentum $\mathbf{p}$ to a state with quasi-momentum $\mathbf{p}'$ has the form [22]:

$$W_{ac}(\mathbf{p},\mathbf{p}') = \frac{2\pi D_a^2 kT}{\hbar L^2 \rho s^2}\delta\big(\varepsilon(\mathbf{p}') - \varepsilon(\mathbf{p})\big), \tag{3}$$

where $D_a$ is the deformation-potential constant, $L$ is the characteristic size, $\rho$ is the surface (two-dimensional) density of the material, $s$ is the sound speed in the material, and $\varepsilon$ is the electron energy spectrum.

An analogous formula for scattering probability for optical phonons has the form:

$$W_{op}(\mathbf{p},\mathbf{p}') = \frac{\pi D_o^2}{L^2 \rho \omega_0}\delta(\varepsilon(\mathbf{p}') - \varepsilon(\mathbf{p}) + \hbar\omega_0), \tag{4}$$

where $D_o$ is the deformation-potential constant and $\omega_0$ is the optical phonon frequency.

The total probability of an electron scattering on a given type of phonon is found by integrating the corresponding probability over all final quasi-momenta $\mathbf{p}'$:

$$W(\mathbf{p}) = \frac{L^2}{(2\pi\hbar)^2}\int W(\mathbf{p},\mathbf{p}')d\mathbf{p}'. \tag{5}$$

Let us find the total probability per unit time of scattering by acoustic and optical phonons in phosphorene. We write the total scattering probability for acoustic phonons by substituting (4) into (6):(3)(5)

$$W_{ac}(\mathbf{p}) = \frac{D_a^2 kT}{2\pi\hbar^3 \rho s^2}\int_{-\infty}^{\infty}\delta\big(\varepsilon(\mathbf{p}') - \varepsilon(\mathbf{p})\big)dp_x dp_y.$$

We use the property of the $\delta$-function:

$$\delta\big(f(x)\big) = \sum_i \frac{\delta(x - x_i)}{|f'(x_i)|}, \tag{6}$$

where $x_i$ are the roots of the equation $f(x_i) = 0$.

To find the roots $x_i$, we denote

$$\varepsilon(\mathbf{p}') = E, \tag{7}$$

where $E$ is a constant. Substituting the energy spectrum (2) into (8):(7)(1)

$$\frac{\Delta {p_x'}^2}{m_x} + \left(\Delta + \frac{{p_y'}^2}{2m_y}\right)^2 = E^2.$$

Hence the roots are:

$$p'_{x\pm} = \pm\sqrt{\frac{m_x}{\Delta}\left(E^2 - \left(\Delta + \frac{{p_y'}^2}{2m_y}\right)^2\right)}.$$

The roots are defined only under condition:

$$E^2 - \left(\Delta + \frac{{p'_y}^2}{2m_y}\right)^2 \geq 0. \tag{8}$$

The derivative of $f(p'_x)$:

$$f'(p'_x) = (\varepsilon(\mathbf{p}') - E)' = \frac{1}{\varepsilon(\mathbf{p}')}\frac{\Delta}{m_x}p'_x.$$

At the points where $\varepsilon(\mathbf{p}') = E$, we obtain:

$$f'(p'_{x\pm}) = \frac{\Delta}{m_x}\frac{p'_{x\pm}}{E}.$$

We rewrite the $\delta$-function in accordance with (7):(6)

$$\delta\big(\varepsilon(\mathbf{p}') - \varepsilon(\mathbf{p})\big) = \Theta\frac{m_x}{\Delta}\frac{E}{|p'_{x+}|}\delta(p'_x - p'_{x+}) + \Theta\frac{m_x}{\Delta}\frac{E}{|p'_{x-}|}\delta(p'_x - p'_{x-}), \tag{9}$$

where $\Theta = \theta\left(E^2 - \left(\Delta + \frac{{p'_y}^2}{2m_y}\right)^2\right)$ is the Heaviside function ensuring condition (9).(8)

The integral of the $\delta$-function (10) over (9)$p'_x$:

$$I_x = \int_{-\infty}^{\infty}\delta\big(\varepsilon(\mathbf{p}') - \varepsilon(\mathbf{p})\big)dp'_x = \Theta\frac{m_x}{\Delta}\int_{-\infty}^{\infty}\frac{E}{|p'_{x+}|}\delta(p'_x - p'_{x+})dp'_x + \int_{-\infty}^{\infty}\frac{E}{|p'_{x-}|}\delta(p'_x - p'_{x-})dp'_x.$$

In view of the properties of the $\delta$-function, and also taking $|p'_{x+}| = |p'_{x-}|$ into account, we obtain:

$$I_x = 2E\sqrt{\frac{m_x}{\Delta}}\frac{\Theta}{\sqrt{E^2 - \left(\Delta + \frac{{p'_y}^2}{2m_y}\right)^2}}.$$

Next we calculate the integral over $p'_y$:

$$I_y = 2\sqrt{\frac{m_x}{\Delta}}E\int_{-\infty}^{\infty}\frac{\Theta}{\sqrt{E^2 - \left(\Delta + \frac{{p'_y}^2}{2m_y}\right)^2}}dp'_y$$

We eliminate the Heaviside function $\Theta$ by changing the integration limits. From the inequality (9) we obtain (8)$p'_y \in \left(-\sqrt{2m_y(E - \Delta)}, \sqrt{2m_y(E - \Delta)}\right)$. Denote $b = \sqrt{2m_y(E - \Delta)}$. Then the integral is rewritten as:

$$I_y = 2\sqrt{\frac{m_x}{\Delta}}E\int_{-b}^{b}\frac{1}{\sqrt{E^2 - \left(\Delta + \frac{{p'_y}^2}{2m_y}\right)^2}}dp'_y.$$

The obtained integral reduces to the complete elliptic integral of the first kind $K(m)$:

$$I_y = 4\sqrt{\frac{m_x}{\Delta}}E\sqrt{\frac{2m_y}{\Delta+E}}K\left(\frac{\Delta-E}{\Delta+E}\right) = 4E\sqrt{\frac{2m_x m_y}{\Delta(\Delta+E)}}K\left(\frac{\Delta-E}{\Delta+E}\right),$$

$$K(m) = \int_0^{\pi/2} \frac{dx}{\sqrt{1-m\sin^2 x}}.$$

Then the total probability per unit time of scattering by acoustic phonons has the form:

$$W_{ac}(\mathbf{p}) = \frac{2D_a^2 kTE}{\pi\hbar^3\rho s^2}\sqrt{\frac{2m_x m_y}{\Delta(\Delta+E)}}K\left(\frac{\Delta-E}{\Delta+E}\right) \tag{10}$$

Let us find the total probability of scattering by optical phonons by substituting (5) into (6):(4)(5)

$$W_{op}(\mathbf{p}) = \frac{D_o^2}{4\pi\hbar^2\rho\omega_0}\int_{-\infty}^{\infty}\delta(\varepsilon(\mathbf{p}')-\varepsilon(\mathbf{p})+\hbar\omega_0)dp_x dp_y$$

The integration of the $\delta$-function is performed as in the case of acoustic phonons, but up to a quantity $E$, which in the case of optical phonons has the form:

$$E_o = \varepsilon(\mathbf{p}) - \hbar\omega_0.$$

Hence we obtain the total probability of scattering by optical phonons:

$$W_{op}(\mathbf{p}) = \frac{D_o^2 E_o}{\pi\hbar^2\rho\omega_0}\sqrt{\frac{2m_x m_y}{\Delta(\Delta+E_o)}}K\left(\frac{\Delta-E_o}{\Delta+E_o}\right). \tag{11}$$

The obtained total scattering probabilities (11) and (12) are used in (3).(10)(11)(2)

### 3. Results and discussion

The following parameters were fixed for the simulation: $L_x = L_y = 300$ nm; $m_x = 1.285m_e$, $m_y = 0.125m_e$ [16]; $T_0 = 300$ K; $\Delta = 1$ eV [23]; $\rho = 1.4 \cdot 10^{-3}$ g/m$^2$, the thickness of the monolayer phosphorene sample $d = 5.3 \cdot 10^{-10}$m [24]. The $X$-axis corresponds to the $ZZ$ (zigzag) axis, the $Y$-axis corresponds to the $AC$ (armchair) axis (Fig. 1). For the single acoustic scattering mechanism the deformation-potential constant was taken as the average over the two directions $2D_a = 4.9$ eV [25]. For the two optical scattering mechanisms (with absorption and emission) the deformation-potential constant $D_o = 1.7 \cdot 10^{11}$ eV/m, the optical and the phonon energy $\hbar\omega_o = 32$ meV [26].

It should be noted that during the simulations the influence of including various combinations of charge carrier scattering mechanisms on the obtained thermal conductivity value was experimentally checked. The check showed that the determining mechanism is the inelastic scattering of charge carriers on optical phonons. Therefore acoustic-phonon scattering is treated in the simplest form, without taking the anisotropy into account.

It should also be noted that only one branch of optical phonons (LO) was chosen, since this branch is the dominant one for scattering [26].

In [14, 15] the following values of the electronic thermal conductivity coefficients were obtained:

$$\kappa_{zz}^{ref} = 3.6\frac{\mathrm{W}}{\mathrm{K}\cdot\mathrm{m}}, \qquad \kappa_{ac}^{ref} = 37\frac{\mathrm{W}}{\mathrm{K}\cdot\mathrm{m}}. \tag{12}$$

At a surface carrier concentration of $n_{2D}^0 = 2\cdot 10^{17}\mathrm{m}^{-2}$ (according to experimental works [27-29], such concentrations are attainable in black phosphorene samples) the Monte Carlo simulation using the Green-Kubo approach gives the following values of the electronic thermal conductivity coefficient:

$$\kappa_{zz} \approx 1.12\ \frac{\mathrm{W}}{\mathrm{K}\cdot\mathrm{m}}, \qquad \kappa_{ac} \approx 46.30\frac{\mathrm{W}}{\mathrm{K}\cdot\mathrm{m}}. \tag{13}$$

The obtained values (14) are of the same orders of magnitude as (13).(13)(12)

When studying thermal conductivity, it is of interest to determine the relative contributions of the electronic and lattice parts. In the work [14, 15] a method for the experimental determination of the electronic thermal conductivity contribution was proposed on the basis of the following considerations. For the simultaneous consideration of the electronic and the phonon parts of the thermal conductivity we write Fourier's law in the following form:

$$\begin{gathered}\boldsymbol{q} = \boldsymbol{q}_{el} + \boldsymbol{q}_{ph} = -\kappa_x^{el}\frac{\partial T}{\partial x}\boldsymbol{i} - -\kappa_y^{el}\frac{\partial T}{\partial y}\boldsymbol{j} - \kappa_x^{ph}\frac{\partial T}{\partial x}\boldsymbol{i} - \kappa_y^{ph}\frac{\partial T}{\partial y}\boldsymbol{j} = \\ = -\left(\kappa_x^{el} + \kappa_x^{ph}\right)\frac{\partial T}{\partial x}\boldsymbol{i} - \left(\kappa_y^{el} + \kappa_y^{ph}\right)\frac{\partial T}{\partial y}\boldsymbol{j} = q_x\boldsymbol{i} + q_y\boldsymbol{j}.\end{gathered} \tag{14}$$

Due to the anisotropy of the thermal conductivity in black phosphorene the heat flux density vector $\boldsymbol{q}$ and the temperature gradient vector $\boldsymbol{\nabla}T$ do not coincide in direction (Fig. 3). The dependence of the angle $6\Phi(\theta)$ (where $\Phi$ is the angle between the heat flux and the temperature gradient and $\theta$ is the angle between the temperature gradient and the $X$-axis) is shown in Fig. 4 for the phonon and electronic heat fluxes separately. This dependence is similar in character to the corresponding dependence in [14, 15]. The dependence $8\Phi(\theta)$ exhibits a non-monotonic behaviour with a maximum at $\theta = \frac{k\pi}{2}, k \in Z$. The difference between the minimal and maximal values of $\Phi$ strongly depends on the material parameters – the concentration of charge carriers $n_{2D}$ and the mean frequency of scattering of charge carriers on lattice inhomogeneities $\nu$ (due to the influence of the electronic part of the thermal conductivity). Changing these parameters and constructing the plots of $\Phi(\theta)$ for different values of the electronic thermal conductivity coefficient $\kappa_{ac,zz}$, we calculate the differences between the minimal and maximal values of $\Phi$ along these plots and then invert the dependence $\kappa_{ac,zz}(\Phi_{max} - \Phi_{min})$. In this way a method for separating the electronic thermal conductivity contribution is obtained that is based on the measuring the angle between the heat flux vector $\boldsymbol{q}$ and the temperature gradient vector $\boldsymbol{\nabla}T$ [14, 15]. In Figs. 5, 6, 7 the dependences 101214$\Phi(\theta)$ for the total heat flux are presented for the concentrations $n_{2D}^0$, $10n_{2D}^0$, $0.1n_{2D}^0$, respectively. It can be seen that they also agree with the corresponding plots in [14, 15].

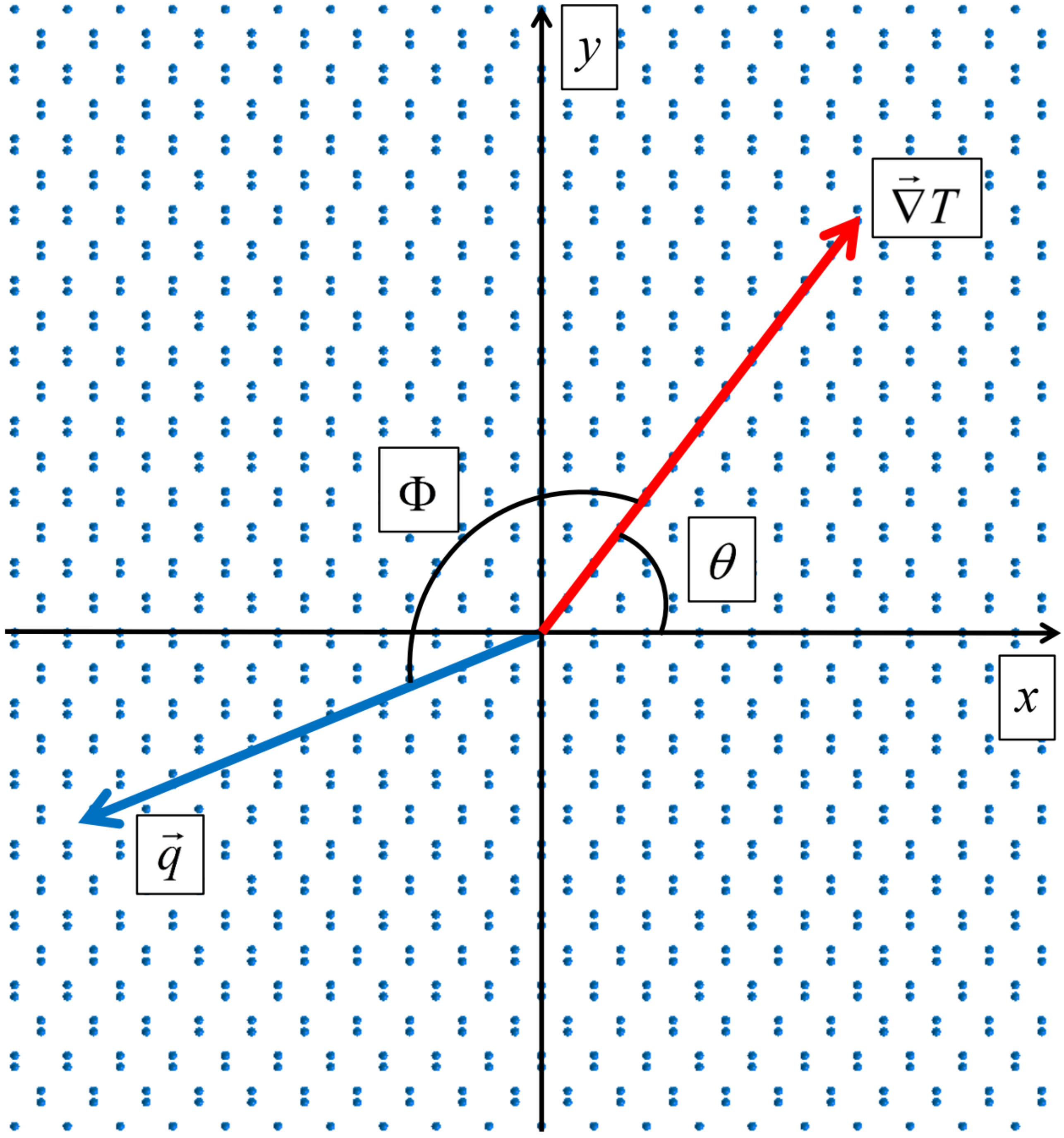

**Fig. 3.3** Problem geometry: $\Phi$ is the angle between the temperature gradient vector $\vec{\nabla}T$ and the heat flux density vector $\vec{q}$; $\theta$ is the angle between the temperature gradient vector $\vec{\nabla}T$ and the $OX$

The Monte Carlo method used in the present work for the calculating the electronic thermal conductivity, by directly taking into account the scattering of charge carriers on optical and acoustic phonons and by the calculation of the correlations between the heat fluxes and the particle fluxes within the Green-Kubo approach, takes into account the microscopic transport mechanisms and is a more general method than the constant momentum relaxation time approximation used in [14, 15]. The close agreement between the electronic thermal conductivity coefficients obtained in the present work and those reported in[14, 15], as well as the qualitative agreement of the plots of the angle $\Phi$ between the heat flux vector $\boldsymbol{q}$ and the temperature gradient vector $\boldsymbol{\nabla}T$ versus the angle $\theta$ between the temperature gradient and the $X$-axis, may serve as a validation of the approach and method for extracting the electronic contribution to the thermal conductivity proposed in [14, 15].

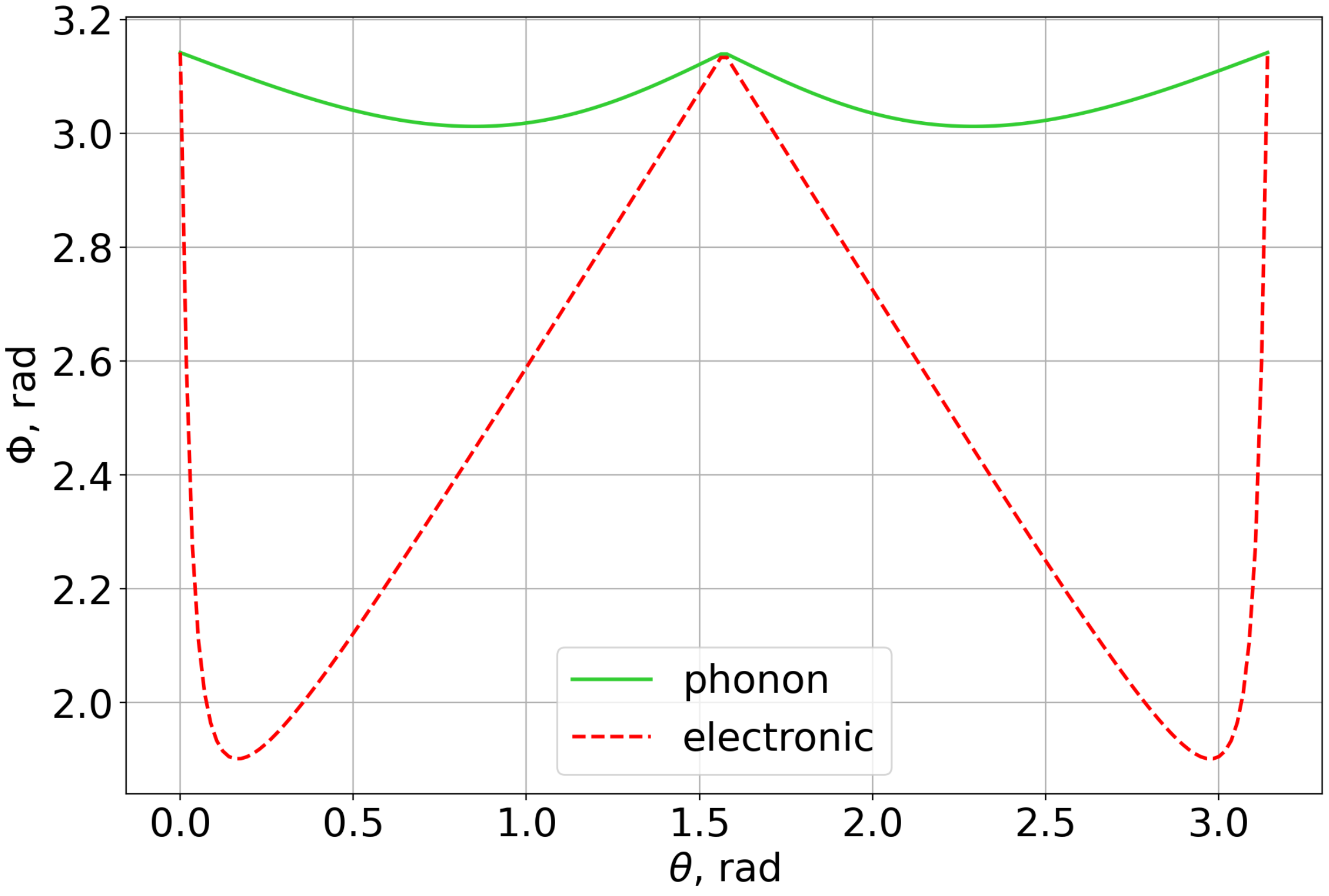

**Fig. 4.4** The $\Phi(\theta)$ dependence for the phonon and the electronic heat fluxes.

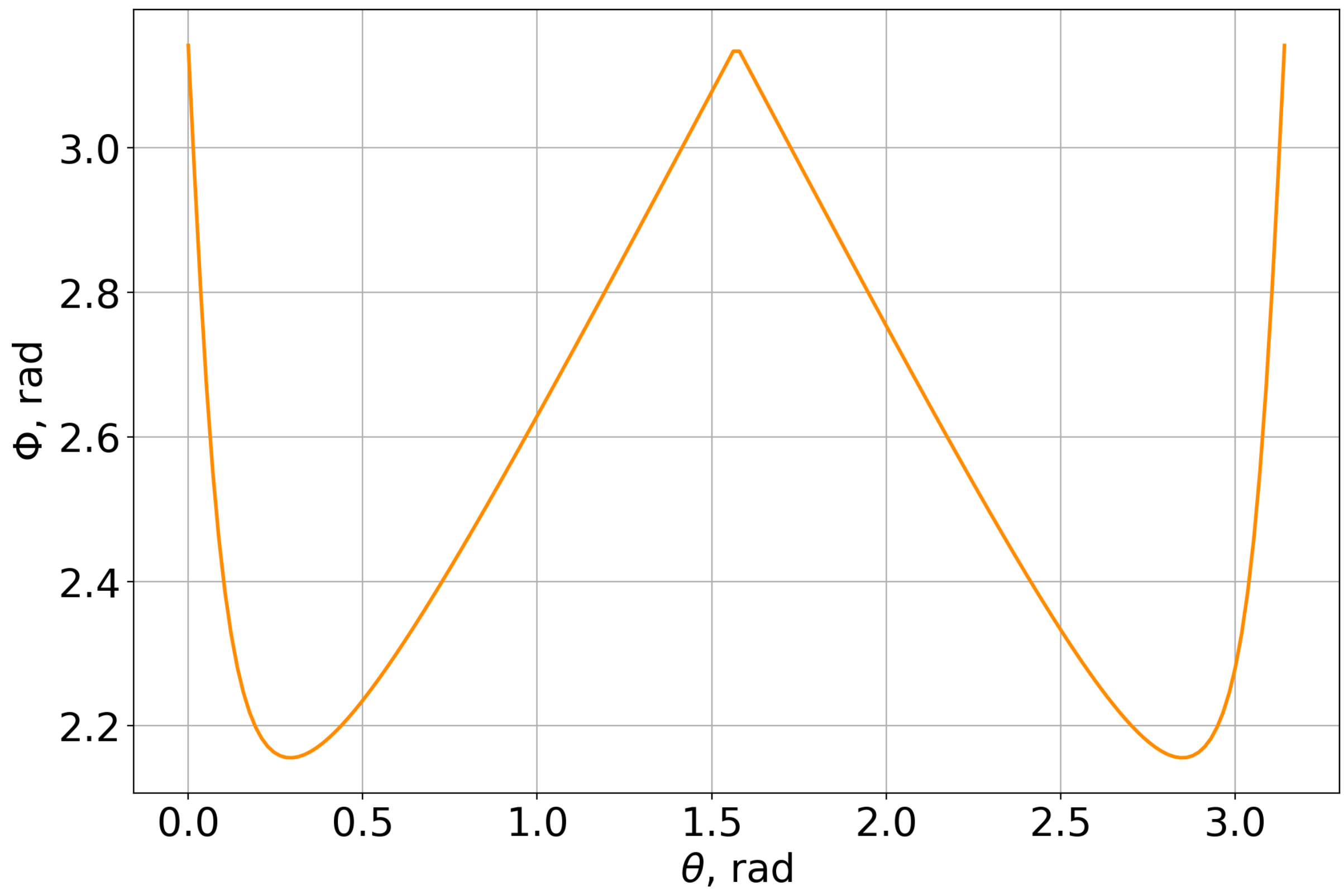

**Fig. 5.5** The $\Phi(\theta)$ dependence for the total heat flux for a charge carrier concentration $n_{2D}^0$

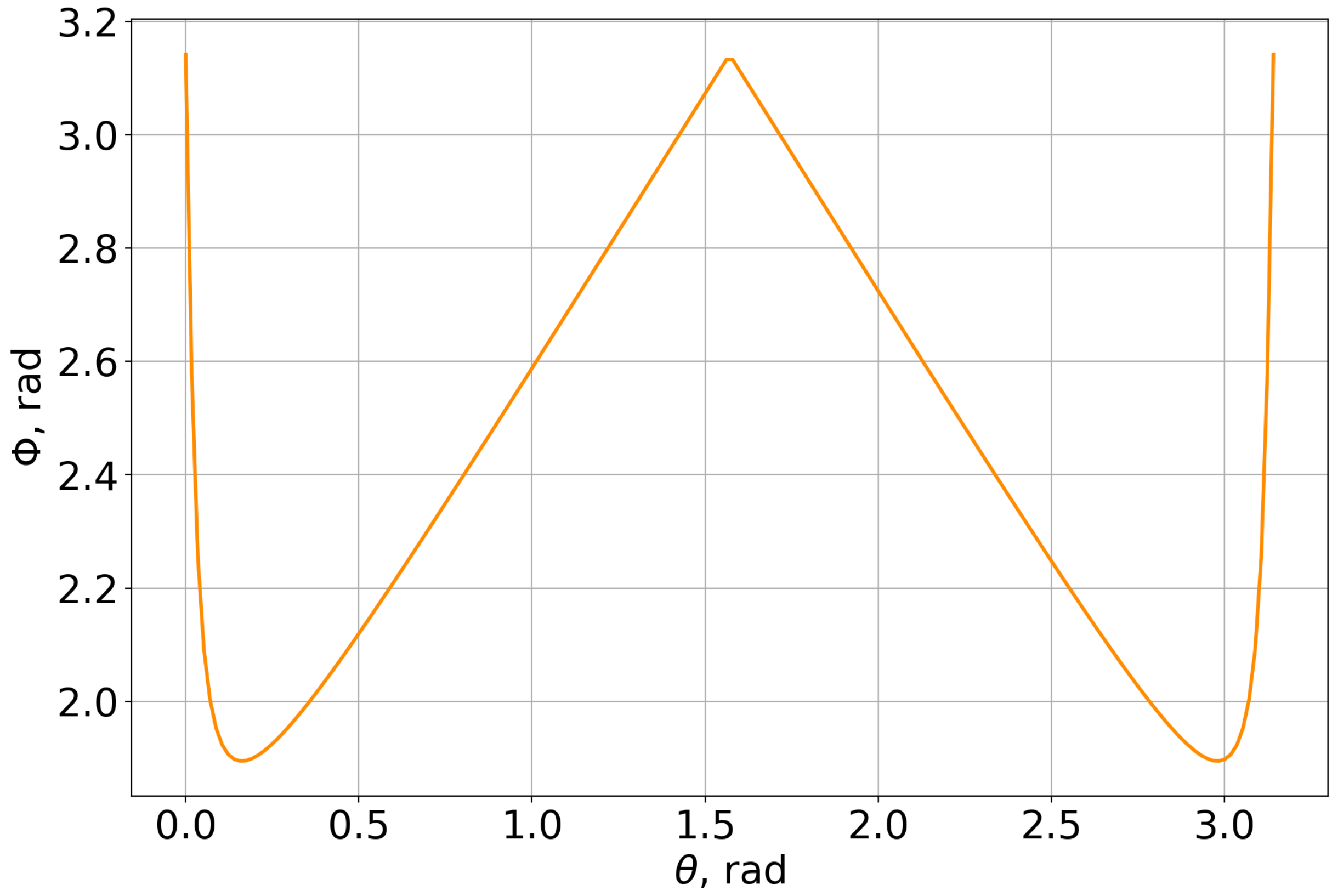

**Fig. 6.6** The $\Phi(\theta)$ dependence for the total heat flux for a charge carrier concentration $10n_{2D}^0$

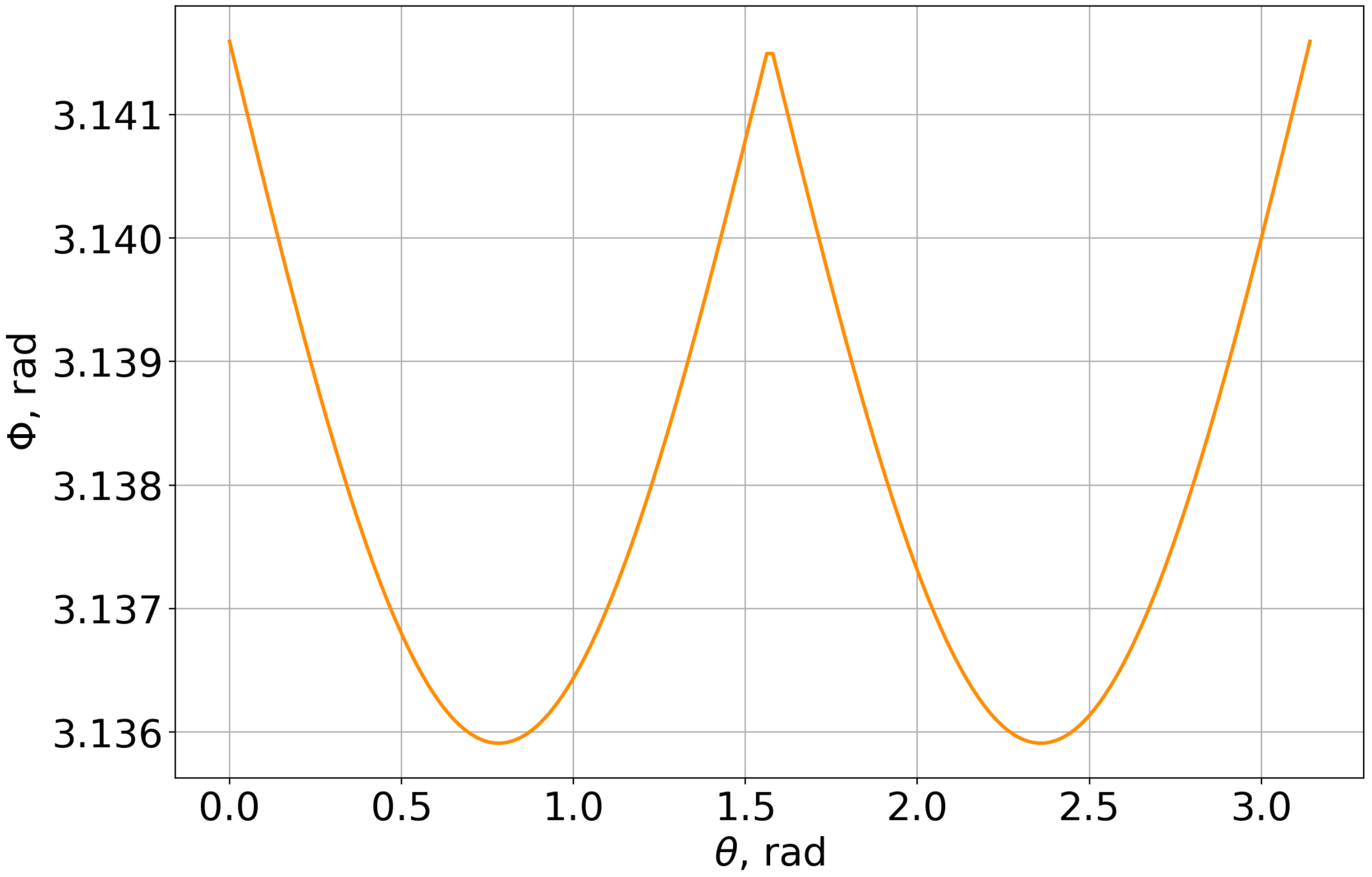

**Fig. 7.7** The $\Phi(\theta)$ dependence for the total heat flux for a charge carrier concentration $0.1n_{2D}^0$

**Conclusions**

As a result of the study, the estimates of the electronic thermal conductivity coefficient of phosphorene were obtained using the ensemble Monte Carlo method within the Green-Kubo approach. The calculated values indicate that at a temperature of 300 K the electronic thermal conductivity along the armchair direction is an order of magnitude higher than along the zigzag direction.

The main result of the work is a more rigorous justification of the results of the study [14, 15]. This strengthens the assumption that experimental measurements of the dependence of the angle $\Phi$ between the heat flux vector $\boldsymbol{q}$ and the temperature gradient vector $\nabla T$ on the angle $\theta$ between the temperature gradient and the $X$-axis may allow to quantitatively extract the electronic contribution to the heat flux and to the thermal conductivity.

**Funding**

This work was not funded by grants from governmental, commercial or non-profit organizations.

**Conflict of Interest**

The authors declare that they have no known financial conflicts of interest or personal relationships that could have influenced the results presented in this article.